\documentclass{article}
\usepackage{spconf,amsmath,graphicx}
\usepackage{xcolor}
\usepackage{cite}
\usepackage{hyperref}
\usepackage{multirow}
\usepackage{todonotes}
\usepackage[tone,extra,safe]{tipa}
\usepackage{booktabs}
\usepackage{amssymb}

\title{How Phonotactics Affect Multilingual and Zero-shot ASR Performance}
\name{\begin{tabular}{c}Siyuan Feng$^{1\dagger}$,
Piotr \.Zelasko$^{2,3\dagger}$, 
Laureano Moro{-}Vel{\'{a}}zquez$^{2}$,
Ali Abavisani$^{4}$,\\
Mark Hasegawa-Johnson$^{4}$, Odette Scharenborg$^{1}$, Najim Dehak$^{2,3}$\end{tabular}
\thanks{$^{\dagger}$Equal contribution. Code available at \url{https://github.com/pzelasko/kaldi/tree/discophone/egs/discophone}.}
}

\address{$^{1}$Multimedia Computing Group, Delft University of Technology, Delft, The Netherlands \\
$^{2}$Center for Language and Speech Processing, $^{3}$Human Language Technology Center of Excellence, \\Johns Hopkins University, Baltimore, MD, USA\\
$^{4}$Department of Electrical and Computer Engineering, University of Illinois at Urbana-Champaign, IL, USA\\
\begin{footnotesize}
\texttt{s.feng@tudelft.nl, piotr.andrzej.zelasko@gmail.com}
\end{footnotesize}
    }
\begin{document}
\ninept
%
\maketitle
\begin{abstract}
The idea of combining multiple languages' recordings to train a single automatic speech recognition (ASR) model brings the promise of the emergence of universal speech representation.
Recently, a Transformer encoder-decoder model has been shown to leverage multilingual data well in IPA transcriptions of languages presented during training.
However, the representations it learned were not successful in zero-shot transfer to unseen languages.
Because that model lacks an explicit factorization of the acoustic model (AM) and language model (LM), it is unclear to what degree the performance suffered from differences in pronunciation or the mismatch in phonotactics.
To gain more insight into the factors limiting zero-shot ASR transfer, we replace the encoder-decoder with a hybrid ASR system consisting of a separate AM and LM.
Then, we perform an extensive evaluation of monolingual, multilingual, and crosslingual (zero-shot) acoustic and language models on a set of 13 phonetically diverse languages.
We show that the gain from modeling crosslingual phonotactics is limited, and imposing a too strong model can hurt the zero-shot transfer.
Furthermore, we find that a multilingual LM hurts a multilingual ASR system's performance, and retaining only the target language's phonotactic data in LM training is preferable.
\end{abstract}

%


\section{Introduction}
\label{sec:intro}

Automatic speech recognition (ASR) technologies have achieved enormous advancement during the past decade \cite{zhang2020Transformer}. 
Typically, a well-trained deep neural network (DNN) based ASR system \cite{ASAPP-ASR} requires hundreds to thousands of hours of transcribed speech data. 
This requirement can be challenging to most of the world's spoken languages that are considered low-resourced  \cite{speech2020scharenborg}. 
Most languages, including ethnic minority languages in most countries outside Europe, 
lack transcribed speech data, digitized texts, and digitized pronunciation lexicons.
Consequently, high-performance ASR systems are only available for a small number of major languages, with English and Mandarin being prominent examples.

To facilitate the development of ASR technologies in low-resource languages, knowledge transfer across languages has been proposed in acoustic modeling \cite{swietojanski2012unsupervised,Huang2013cross,li2020universal}. 
In \cite{swietojanski2012unsupervised}, a DNN acoustic model (AM) is pretrained with the speech from non-target languages and further retrained with a small amount of speech from the target language. In \cite{Huang2013cross}, a multilingual DNN is trained by simultaneously adopting speech of target and non-target languages. 
A very recent study \cite{li2020universal} proposed an approach to building a nearly-universal phone recognition model. By employing $11$ resource-rich languages in multilingual training, the model could achieve $17\%$ phone error rate (PER) improvements in the low-resource languages studied, compared to a system without using such a nearly-universal phone model.
These results indicate that speech resources from other languages benefit acoustic modeling for a target low-resource language.

Presumably, the advantage of multilingual acoustic modeling 
can be explained by the possibility of sharing phonetic representations between languages.
Most past research reports performance improvements from monolingual to multilingual training in terms of overall PER or word error rate (WER).
It is not clear, however, what aspects of phonetic representations a multilingual AM is learning and 
whether the benefit of adopting multilingual acoustic modeling is consistent
for every phone.
To gain a deeper understanding of how phonetic representations are shared and transferred across languages, our recent work \cite{Zelasko2020That} adopted the end-to-end (E2E) ASR architecture \cite{kim2017joint} to perform International Phonetic Alphabet (IPA)  token sequence recognition tasks.
Extensive analyses of the results revealed that all phone tokens, including language-unique ones, can benefit from multilingual training. 
Analyses also showed that the representations learned in \cite{Zelasko2020That} were less effective in transferring to unseen languages.

The question is to what degree the ASR performance degradation to unseen languages~\cite{Zelasko2020That}   suffered from differences in pronunciation or the mismatch in phonotactics\footnote{Phonotactics defines restrictions in a language on the permissible combinations of phonemes.}.
Here, we  use the hybrid ASR system to investigate the role of language-specific  phonotactic 
information on phone recognition performance, especially under a \textbf{zero-shot} scenario (which we refer to as crosslingual ASR in this paper) where speech and linguistic resources are unknown for a target language. 
In an encoder-decoder ASR system, phonotactic and acoustic information are learned within a single model, making it 
difficult to separate the effects of phonotactic and acoustic information; a hybrid ASR, by contrast,
makes such experiments natural and effective.


In the present study, we strive to answer the above question by adopting the hybrid DNN-hidden Markov model (DNN-HMM) ASR architecture to conduct a series of IPA phone recognition experiments. 
In our hybrid ASR systems, we model basic acoustic units or phones. We will use a phone categorization assuming that a phone can be represented by one or several IPA symbols (letters and modifiers), corresponding to a single phone. We define these phones as \textbf{IPA phones}. For instance, [\textipa{a}] and [\textipa{a}\textlengthmark\tone{5}\tone{1}] will be considered as two different IPA phones, independently of the language. This is different from \cite{Zelasko2020That}, in which each IPA symbol (including modifiers, e.g., the tone symbols) was modeled as a base unit in the E2E ASR systems.
To measure the effect of phonotactic information, phonotactic language models (LMs) of different capacities are trained to decode alongside the hybrid DNN AM. Besides, a word LM which could impose a much stronger phonotactic constraint, is trained and compared with the phonotactic LMs. 
The experiments are carried out with $13$ phonetically diverse languages, based on the selection criteria in \cite{Zelasko2020That}.  
Three ASR training scenarios are considered: monolingual as a baseline; multilingual, to investigate the effect of merging data from all the $13$ languages; and crosslingual, namely a zero-shot learning task,
to investigate whether universal phone representations exist that can be used for recognition in an unknown language. 
\section{Experimental setup}
\begin{table}[!t]
\renewcommand\arraystretch{0.7}
\centering
\caption{Speech data used in the experiments. Train and Eval columns are presented in the number of hours. Vow and Con stand for the number of distinct IPA phones used to describe vowels and consonants, respectively. Uniq denotes the number of IPA phones only existing in that language. The symbol $^{\ddagger}$ indicates tone languages.}
\resizebox{ 0.99 \linewidth}{!}{%
\begin{tabular}{c|l|ccccc}      
\toprule
Corpus & Language & Train & Eval & Vow & Con & Uniq \\
\midrule
\multirow{5}{*}{GlobalPhone} &
Czech & 24.0 & 3.8 & 10 & 25 &5\\
&French & 22.8 &  2.0 &19&26&9\\
&Spanish & 11.5 &  1.2 &6&24&4\\
&Mandarin$^{\ddagger}$ & 14.9 & 1.6&60&33&41\\
&Thai$^{\ddagger}$ & 22.9 & 0.4&108&35&17\\
\midrule
\multirow{8}{*}{Babel} &
Cantonese$^{\ddagger}$ &126.6 & 17.7 &110 &25&88\\
&Bengali & 54.5 & 9.8&19&33&13\\
&Vietnamese$^{\ddagger}$ & 78.2 & 10.9&204 &25&130\\
&Lao$^{\ddagger}$ & 58.7 & 10.5&106&61&83\\
&Zulu$^{\ddagger}$ & 54.4 & 10.4&15&54&19\\
&Amharic &  38.8 & 11.6 &8& 52&9\\
&Javanese & 40.6 & 11.3 &11&23&1\\
&Georgian &  45.3 & 12.3 &5&30&2\\
\bottomrule
\end{tabular}%

}
\label{tab:database}
\end{table}
\subsection{Databases}
There are in total $13$ languages chosen in this study to mimic the setup of~\cite{Zelasko2020That}, as shown in Table \ref{tab:database}. Table \ref{tab:database} reports the number of distinct IPA phones \cite{international1999handbook} used to describe the phone inventories of each language. 
Note that for tone languages, phonemes that differ only in tone levels are described by distinct IPA phones. For instance, the high-, rising-, dipping- and falling-tone  /i/ in Mandarin are described by four different IPA phones [i\tone{55}], [i\tone{33}\tone{55}], [i\tone{22}\tone{11}\tone{44}], and [i\tone{55}\tone{11}], respectively.   

Speech recordings of the $13$ languages are taken from two corpora: GlobalPhone \cite{Schultz02globalphone} and IARPA Babel.
GlobalPhone has a smaller number of recordings and is characterized by limited noise and reverberation. Babel contains more hours of training data and is significantly more challenging to recognize due to more naturalistic recording conditions. We follow train, development, and evaluation data splits adopted in our previous work \cite{Zelasko2020That}. Note that the high number of vowels in tone languages is due to different combinations of tone modifiers (e.g., Vietnamese and Cantonese each have six tones per vowel).





\subsection{Implementation of hybrid ASR systems}
We adopt the hybrid DNN-HMM architecture~\cite{dahl2012context} for training the IPA phone recognition systems in this study. 
To obtain the IPA phonetic transcriptions for ASR system training and evaluation, LanguageNet G2P models~\cite{hasegawa2020grapheme} are leveraged to convert orthographic transcriptions in the GlobalPhone and Babel corpora into IPA phonetic sequences. 

The hybrid ASR AM is implemented using the Kaldi toolkit~\cite{povey2011kaldi}. 
It is a factorized time-delay neural network (TDNNF) AM~\cite{povey2018semi}, consisting of $12$ layers, with a hidden dimension of $1024$ and bottleneck dimension of $128$ and Resnet-style skip connections. The TDNNF AM is trained with the lattice-free maximum mutual information (LF-MMI) criterion~\cite{povey2016purely} for $4$ epochs. Frame-level phone alignments used as supervision for the TDNNF model training are obtained by forced-alignment with a GMM-HMM AM trained beforehand. The input features consist of $40$-dimension high-resolution MFCCs, $3$-dimension pitch features \cite{ghahremani2014pitch} and $100$-dimension i-vectors. The $43$-dimension MFCC+pitch feature and its left and right neighboring feature are appended and further complemented with i-vectors to form the TDNNF AM input. 
In TDNNF training, we use the hyperparameters from Kaldi's Wall Street Journal recipe\footnote{\texttt{wsj/s5/local/chain/tuning/run\_tdnn\_1g.sh}} without further tuning in our experiments. An identical TDNNF structure is used for each of the experiments in this study.

One possible concern about the validity of our approach is that a sequence-level criterion (e.g. LF-MMI) might "leak" the language model into the acoustic model. In principle, this is possible -- however, we believe that LF-MMI is less likely to cause that (as compared to e.g. CTC) since it imposes a (weak) language model supervision during training, allowing the model to spend less effort trying to learn it. In either case, the acoustic model network’s frame-wise predictions are conditionally independent of each other, limiting the opportunities for learning a language model.

\begin{table*}[!ht]
\renewcommand\arraystretch{0.6}
\centering
\caption{PER ($\%$) results for the $13$ languages under Mono(lingual), Cross(lingual) and Multi(lingual) experimental scenarios. Phonotactic uni-gram, bi-gram,  tri-gram LMs, and word tri-gram LM are denoted as -ug, -bg, -tg and -wtg, respectively;
Mono-tg LMs are tested in all experimental settings. Asterisk denotes zero-shot in a weaker sense. The symbol $^{\ddagger}$ indicates tone languages.}
\resizebox{ 1 \linewidth}{!}{%
\begin{tabular}{l|ccc|c||ccc|c|c||ccc|c|cc}      
\toprule
 AM & \multicolumn{4}{c||}{Monolingual} & \multicolumn{5}{c||}{Crosslingual} & \multicolumn{6}{c}{Multilingual}   \\
LM  & Mono-ug & Mono-bg & Mono-tg & Mono-wtg & Cross-ug & Cross-bg & Cross-tg  & Mono-tg$^{*}$  &Cross-wtg & Multi-ug& Multi-bg  &Multi-tg & Mono-tg  & Multi-wtg & Mono-wtg \\
\midrule
Czech&18.7&16.1&14.1&\textbf{8.6}&68.8&69.0 &70.5 &\textbf{55.2}&78.2&28.8&25.0& 20.1& 18.6&11.0& \textbf{9.5}\\
French&22.5&20.3&16.7&\textbf{13.0}&63.1&62.9&64.1&\textbf{55.1}&76.4&27.2&25.4&21.8&19.9&14.4&\textbf{14.3}\\
Spanish&12.2&11.8&\textbf{10.3}&10.7&56.5&57.2&64.6&\textbf{33.9}&61.4&14.5&12.7&11.1& 11.2 &\textbf{11.0}&11.2\\
Mandarin$^{\ddagger}$&26.9&20.2&\textbf{17.6}&-&87.8&88.2&89.8&\textbf{80.4}&-&43.4&26.1 &\textbf{20.4}&21.0&-&-\\
Thai$^{\ddagger}$&21.9&20.0&\textbf{18.8}&-&83.3&86.8&88.7&\textbf{73.6}&-&40.2&32.2 &\textbf{27.2}&27.5&-&-\\
\midrule
Cantonese$^{\ddagger}$&61.2&44.7&39.8&\textbf{34.8}&91.2&90.8&91.0&\textbf{87.6}&90.8&75.9&55.2&43.6&42.9&39.1&\textbf{38.8}\\
Bengali&59.8&46.9&41.1&\textbf{41.0}&86.0&83.1&81.3&\textbf{66.8}&87.7&67.9&56.1 &45.6&\textbf{43.3}&44.5&43.7\\
Vietnamese$^{\ddagger}$&54.2&45.0&\textbf{41.6}&42.6&92.5&91.5&91.1&\textbf{87.5}&94.4&67.8&51.3&\textbf{45.8}&\textbf{45.8}&47.7&46.8\\
Lao$^{\ddagger}$&66.6&46.4&\textbf{39.8}&41.8&87.2&\textbf{85.9}&86.0&86.5&92.7&72.4&56.6 &45.1&\textbf{43.3}&48.3&46.2\\
Zulu$^{\ddagger}$&70.6&54.1&44.6&\textbf{40.5}&86.3&83.5&82.3&\textbf{72.8}&86.0&72.2&59.7&47.9&46.1&45.3&\textbf{44.4}\\
Amharic&61.2&49.3&41.6&\textbf{32.7}&87.2&83.4&81.0&\textbf{74.8}&82.2&72.7&61.1 &50.2&45.7&39.2&\textbf{38.2}\\
Javanese&57.6&52.3&48.3&\textbf{45.3}&87.0&84.6&85.0&\textbf{69.6}&87.4&65.1&60.9 &56.8&50.6&53.3&\textbf{49.9}\\
Georgian&52.8&42.9&39.1&\textbf{34.6}&85.5&81.4&80.0&\textbf{63.9}&86.8&60.4&48.4 &42.4&39.9&38.8&\textbf{37.3}\\
\bottomrule
\end{tabular}%
}
\label{tab:PER_results}
\end{table*}

The LM in the hybrid ASR system is an n-gram phonotactic LM estimated using the SRILM toolkit \cite{Stolcke02srilm--}. 
IPA phonetic transcripts converted from the GlobalPhone and Babel training graphemic transcripts are used to train the phonotactic LM.
We implement uni-gram, bi-gram, and tri-gram LMs and compare their performances in scoring the TDNNF AM to explore the effect of imposing different amounts of phonotactic information on a hybrid ASR system in recognizing IPA phone sequences. We also experiment with decoding using a word-level tri-gram LM and converting the word lattices to phone lattices before scoring. This way we can impose a much stronger phonotactic constraint on the ASR model than would be possible with phone-level n-grams.
 


There are three experimental scenarios considered in this work.
In the monolingual scenario, a TDNNF AM and an LM are trained using data from one language, and are tested on the same language's evaluation set. 
In the crosslingual (zero-shot) scenario, for each experiment, one language that is going to be tested is taken out from training, and data for the remaining $12$ languages are merged to train a \textit{crosslingual AM} and a \textit{crosslingual LM}. We also test scoring the crosslingual AM using a monolingual LM trained with the left-out language's data.
In this case, it is not zero-shot transfer anymore, as the monolingual LM encodes the phonotactic information of the target language. It allows us to remove the phonotactics from the equation and check the transferability of the AM alone.
In the multilingual scenario, all the $13$ languages' training data are merged to train a multilingual TDNNF AM, which is tested with the evaluation sets of each of the $13$ languages. 
There are two types of LMs used to score the multilingual TDNNF AM: one multilingual LM, which is trained with data from all the $13$ languages, and one monolingual LM, which is trained only using the evaluation language. The use of the latter LM allows us to measure the effect of phonotactics on multilingual ASR performance.


\section{Experimental results}

Experimental results of the IPA phone recognition experiments
under monolingual, crosslingual, and multilingual scenarios are listed in Table \ref{tab:PER_results}\footnote{We did not obtain results using word LMs for Mandarin and Thai because their transcripts in GlobalPhone are not tokenized. Hence, without additional preprocessing, a ``word-level’' LM constructed for them would have used sentences as atomic symbols.}. These results are reported in phone error rate (PER)\footnote{The results reported in this work are not directly comparable to our recent work \cite{Zelasko2020That}, due to different evaluation metrics (PER and phonetic token error rate (PTER)).}.



 \subsection{Comparison of acoustic models}
 Table \ref{tab:PER_results} suggests that the crosslingual TDNNF  AM performs much worse than the monolingual and multilingual TDNNF AMs. When scored with the same monolingual tri-gram phonotactic LM (Mono-tg), the crosslingual TDNNF leads to absolute PER degradations ranging from $20\%$ to $65\%$ in the $13$ languages, compared to the monolingual TDNNF. This degradation is caused by the lack of acoustic data for the target language.
 Mandarin, Thai, Cantonese, Vietnamese, and Lao degrade more from monolingual to crosslingual TDNNF AMs (all over $45\%$) than the other languages.
 One possible explanation is that the numbers of language-unique phones  are much larger in tone languages than in non-tone languages (see Table \ref{tab:database}).
 
 Table \ref{tab:PER_results} shows that the
 monolingual TDNNF performs better than the multilingual TDNNF. 
 When scored with mono-tg and with mono-wtg, the multilingual TDNNF consistently underperforms monolingual TDNNF on all the languages evaluated.
 Interestingly, this finding is opposite to that reported in~\cite{Zelasko2020That}. There might be several reasons for this result. First, the number of shared phones across languages: over $800$ distinct IPA phones are modeled in our multilingual TDNNF AM, while in~\cite{Zelasko2020That}, IPA symbols were modeled as base units, and the total number of distinct IPA symbols was only around $100$. These factors prevent our multilingual TDNNF AM from learning as much phonetic representation sharing, as was done in \cite{Zelasko2020That}. Furthermore, the TDNNF AM output layer is trained to predict the conditional probability distribution of up to $4,000$ senones, learned by a phonetic state-tying decision tree~\cite{young1994tree} alongside the GMM-HMM system used to produce the alignment supervisions. It is not clear whether this benefits or harms the network's ability to share representations across languages. By contrast, the Transformer model in~\cite{Zelasko2020That} predicts IPA symbols directly.
 
 \subsection{The effect of phonotactics}
The effect of imposing different amounts of phonotactic information in hybrid phone-level ASR  is shown in Table \ref{tab:PER_results}. 
We first compare different LMs of tri-gram.
It can be observed that when the multilingual TDNNF is adopted as the AM, scoring with the tri-gram multilingual phonotactic LM (multi-tg) achieves worse -- or at least not better -- performance than the monolingual LM counterpart (mono-tg) in $10$ of the $13$ languages. There is a similar pattern in the word-level LM experiments.
This shows that the phonotactic information from other languages is harmful for phone recognition in a target language with a hybrid ASR system. 

Results in Table \ref{tab:PER_results} suggest that scoring the crosslingual TDNNF AM with mono-tg resulted in PER reductions in $12$ of  the $13$ languages, compared to scoring the same AM but with cross-tg. 
This improvement is gained by imposing phonotactics knowledge of the target language.
This observation demonstrates the importance of phonotactics in the phone-level ASR task and reflects a significant limitation of zero-shot phone recognition: we would never have known the phonotactics of a previously unstudied language. 

We further compare phonotactic LMs of different n-gram orders and the word LM in scoring the hybrid TDNNF AM.
For the monolingual and multilingual TDNNF AMs, the tri-gram phonotactic model outperforms the bi-gram and uni-gram phonotactic models. The word LMs (mono-wtg and multi-wtg) outperform the phonotactic LMs for $8$ out of $11$ languages for the monolingual and multilingual TDNNF AMs. 
As expected, the above observations show that the hybrid phone-level ASR system benefits from imposing more substantial phonotactic constraints of the target language (word LMs comparing to phonotactic LMs) during decoding.
In contrast, in the crosslingual scenario (i.e., when the phonotactic LM does not contain phonotactics of the target language), scoring with cross-wtg is worse than that with cross-tg in all the languages, except for Spanish and Cantonese.
This shows that too strong phonotactic constraints (cross-wtg comparing to cross-tg) from other languages applied to the target language decoding are harmful to phone recognition.
On the other hand, there are benefits to be gained from leveraging some degree of phonotactic knowledge from the non-target languages, as shown by the \emph{cross-tg} performance compared to \emph{cross-ug} and \emph{cross-bg}.
Overall, to infer phone inventories for an unwritten language, leveraging non-target languages to build an LM is desirable at the phone level rather than at the word level.

It is worth noting that when scoring the hybrid ASR systems, we searched for the LM weights that minimize the PER in the range of $2 \sim 17$, which is a broader range than the typical Kaldi setup ($7 \sim 17$) to give the system more freedom in down-weighting the LM if it is not useful. The most frequently chosen weights were $6\sim7$ for the word LMs and $4\sim5$ for the phonotactic LMs, suggesting that it is beneficial to give more emphasis to the acoustic model when the phonotactics are uncertain.

\section{Discussion}

\begin{table}[t]
\renewcommand\arraystretch{0.8}
    \centering
     \resizebox{1\linewidth}{!}
      {
    \begin{tabular}{ccrrrr}
    \toprule
    AM    & LM        & PER & Insertions & Deletions & Substitutions \\
    \midrule
    Mono  & mono-wtg  & 30.0 & 17.6 & 34.0 & 48.4 \\
    Mono  & mono-tg   & 37.7 & 11.5 & 39.6 & 48.9 \\
    Multi & multi-wtg & 34.2 & 18.3 & 29.2 & 52.5 \\
    Multi & multi-tg  & 43.0 &  9.6 & 40.2 & 50.3 \\
    Cross & cross-wtg & 85.1 &  3.2 & 40.6 & 56.1 \\
    Cross & cross-tg  & 82.6 &  2.3 & 42.1 & 55.6 \\
    \bottomrule
    \end{tabular}
    }
    \caption{Phone error rates ($\%$) computed across all languages for mono(lingual), multi(lingual), and cross(lingual) systems and their breakdown into three error type.}
    \label{tab:error_fraction}
\end{table}

We present the errors' breakdown in terms of the balance between the insertions, deletions, and substitutions in Table~\ref{tab:error_fraction}.
We observe that substitutions dominate across all experiments with a contribution of around 50\%, closely followed by deletions. 
Imposing a strong phonotactic prior with a word tri-gram LM (\emph{wtg}) enhances the system performance when target language text data is available for training (21\% relative improvement in \emph{mono} and \emph{multi}) and shifts the error types from deletions towards insertions.
The same is not valid for the zero-shot system - the lack of target languages' phonotactics causes the stronger LM (\emph{cross-wtg}) to fail.

The large number of deletions in ASR is generally concerning, and there may be multiple factors behind it. Firstly, we observe from the phonotactic LM vs.~word LM experiment that a weak phonotactic prior is a significant contributor (see Table~\ref{tab:error_fraction}). 
Secondly, we suspect another factor - both GlobalPhone and Babel transcripts contain a $\langle silence \rangle$ pseudo-word, which sometimes occurs in the middle of an utterance. It is possible that as the alignments are re-estimated in each GMM training step, the silence model is learned using some of the speech regions. 
In our preliminary experiments, silence frames were removed before training the AM, using a re-segmentation strategy based on alignments, however we found out it did not consistently improve the phone recognition performance (not reported in this paper).
Finally, using data augmentation in training would likely help as well, increasing overall AM robustness. However, our preliminary experiments with 3-way speed perturbation \cite{ko2015audio} have not shown a significant change in the crosslingual systems' PER.

A preliminary analysis reveals that the IPA modifier symbol errors account for a noticeable portion of PER in the crosslingual scenario. In other words, the zero-shot ASR system tends to recognize vowels correctly but gets their suprasegmentals and tones wrong. In fact, we noticed that many modifiers are simply deleted, and the system outputs only the base vowel symbol. To investigate the scope of this phenomenon, we scored the system separately with all the IPA modifiers stripped from the reference and hypothesis transcripts. The \emph{cross-tg} system, which scores 82.6\% PER aggregated over all languages, achieves 75.2\% PER in such a lenient scoring scheme. It suggests that disposing of the modifier symbols could help discover languages' base phone inventories with a zero-shot ASR model.

\begin{figure}
    \centering
    \includegraphics[width=\linewidth]{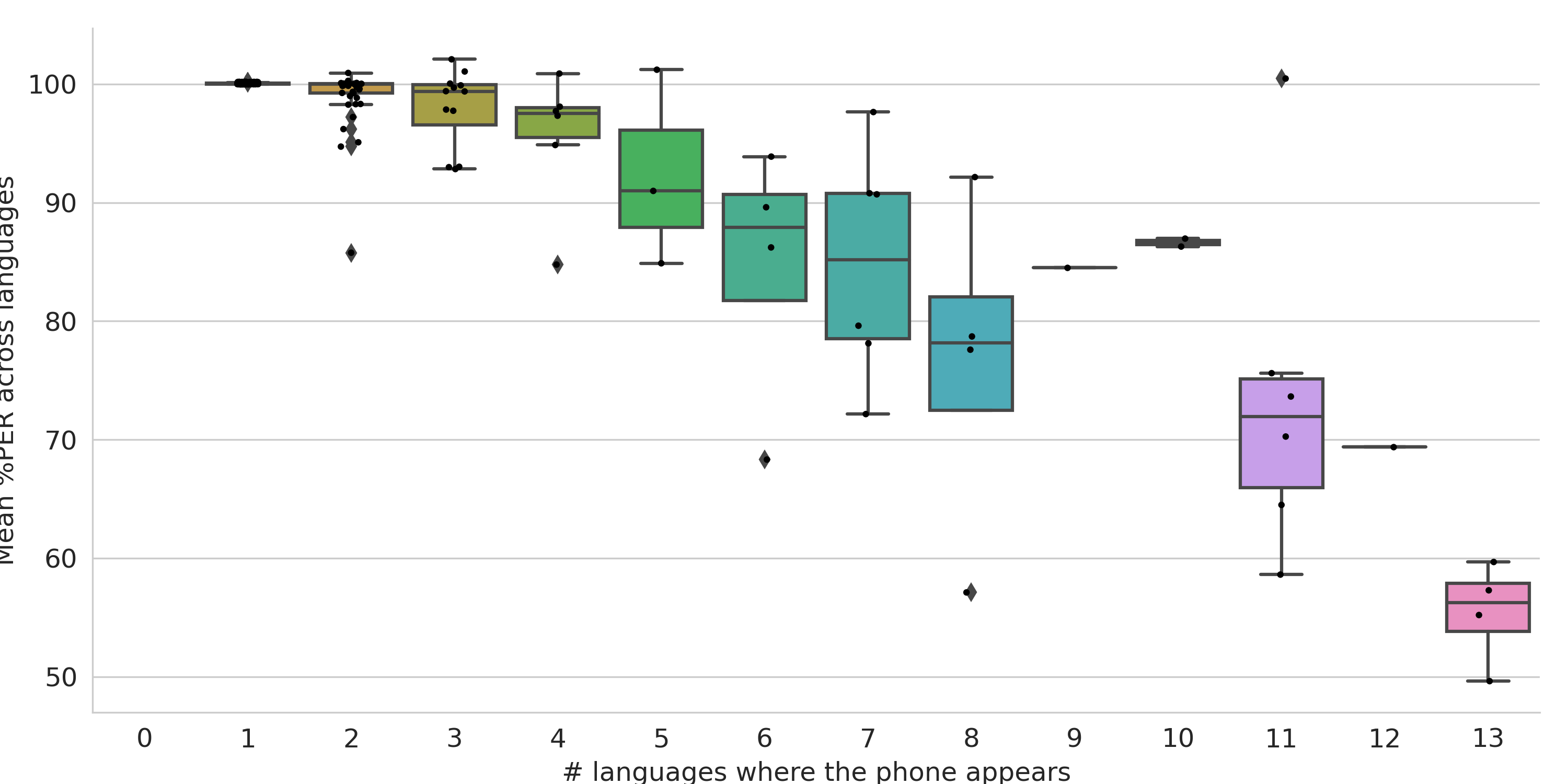}
    \caption{Distribution of PER for different phones, depending on how many languages share that phone, in the crosslingual experiments.}
    \label{fig:crosslingual_per_lang_dist}
\end{figure}

The boxplots in Figure~\ref{fig:crosslingual_per_lang_dist} show individual phones' error rates in the crosslingual experiments, depending on how many languages share a given phone. For language-unique phones, it is expected that their PER will always be 100\%. However, it seems that most of the phones shared by up to 4 languages are also poorly generalized by the ASR system, with almost 100\% PER. While most phones shared between 5-10 languages suffer from terrible performance, there is a stark PER decrease for most phones shared by at least 11 languages. This asserts the findings of~\cite{Zelasko2020That} that there is some degree of representation universality for the phones shared by many languages.

While the comparison of the hybrid ASR performance with that of a Transformer encoder-decoder ASR used in~\cite{Zelasko2020That} is not the primary goal of this work, the setup is similar enough to warrant a brief discussion. Notably, the error rate numbers are not directly comparable, as the Transformer setup used phone tokens as the base units for both prediction and scoring. However, we see that the hybrid monolingual system performance (especially with word LM) is much stronger than that of the Transformer; on the other hand, the Transformer was able to leverage the multilingual data to gain massive performance improvements, which is not true for the hybrid system. It is difficult to say whether the Transformer multi-system outperformed the hybrid mono systems because of the scoring differences and close error rate values. However, it is clear that the Transformer managed to somehow separate the influence of the non-target languages' phonotactics. 
Neither system performed well in the zero-shot scenario.

\section{Conclusions}

We hypothesized that the inclusion of non-target languages' phonotactics in training hurts multilingual and crosslingual ASR systems' performance. We supported that hypothesis with an extensive evaluation of mono-, multi-, and cross- lingual hybrid ASR systems that use either target-only (mono) or joint (multi and cross) phonotactic LMs of different strength (phonotactic uni-, bi-, tri- grams or word tri-grams). We show that while the mono and multi AM systems benefit from a stronger phonotactic constraint, the cross system performs best with a weak phonotactic prior. We also assert that a system's performance with multi AM degrades when combined with multi, instead of mono, LM. We concluded that zero-shot ASR benefits from weakening the phonotactic constraints, which is easily achieved in a hybrid ASR system; it is not yet clear how to similarly weaken phonotactic constraints in an E2E model that learns acoustics and phonotactics jointly. Furthermore, our findings clearly show the deficiencies of zero-shot AM representation transfer, and demonstrate that shared phones suffer less crosslingually. 


\bibliographystyle{IEEEtran}
\begin{small}
\bibliography{refs,strings}
\end{small}

\end{document}